# Multipartite entanglement and Grover's search algorithm

Ri Qu, Juan Wang, Zong-shang Li, Su-li Zhao, Yan-ru Bao, Xiao-chun Cao

*School of Computer Science & Technology, Tianjin University, Tianjin 300072, China*

**Abstract**

We firstly investigate the multipartite entanglement features of the quantum states, including the iteration states achieved by repeated application of Grover iteration and the Oracle ones into which the above iteration states evolve by applying single Oracle operation, employed in Grover's search algorithm by means of the separable degree and the entanglement measure. Then we give the quantitative and qualitative descriptions of the entanglement dynamics in Grover's search algorithm. Our results show that for most instances (i) the separable degrees of these states and ranges of their maximum Schmidt numbers are invariable by following the dynamics of Grover's search algorithm; (ii) the dynamics of Grover's search algorithm is almost "filled" by the fully entangled states.

## I. Introduction

A celebrated result in quantum computation [1] is the discovery of some quantum algorithms [2-5] able to solve problems faster than any known classical algorithm. One of such algorithms is Grover's search algorithm [5] where a quadratic speedup over the best classical one is shown. However, what makes quantum computers powerful is yet not clear. But it is generally believed that quantum entanglement [6,7] plays a key role. Recently, Bruß and Macchiavello [8] have elucidated the role of multipartite entanglement [9] in some quantum algorithms including Grover's search algorithm by studying the entanglement properties of the so-called real equally weighted states (REWS's). They point out that multipartite entanglement is an important property in Grover's search algorithm. Moreover, the paper [10] has qualitatively and completely investigated the entanglement properties of all $n$-qubit REWS's by using the separable and similar degrees. However, most of current researches only consider the entanglement properties of the states achieved by applying one Oracle operation to the uniform superposition state. In this paper, we consider more states that occur by repeated application of Grover iteration or Oracle operation, and analyze their multipartite entanglement properties. Then we give the quantitative and qualitative descriptions of the entanglement dynamics in Grover's search algorithm.

In Grover's search algorithm, the $n$-qubit states that occur after consecutively applying $k$ Grover iterations $G$ are of the form

$$\left|\psi_k^G\right\rangle \equiv \frac{\cos[(2k+1)\theta/2]}{\sqrt{2^n}\cos(\theta/2)} \sum_{x \in f^{-1}(0)} |x\rangle + \frac{\sin[(2k+1)\theta/2]}{\sqrt{2^n}\sin(\theta/2)} \sum_{x \in f^{-1}(1)} |x\rangle, \qquad (1)$$

where $|x\rangle$ represent the computational basis states of $n$ qubits, $f(x)$ is the $\{0,1\}^n \to \{0,1\}$ Boolean function that needs to be evaluated (Note that $f(x)=1$ if and only if $x$ is one of solutions of the search problem) and $\cos(\theta/2) = \sqrt{(2^n - |f^{-1}(1)|)/2^n}$ with $\theta/2 \in [0, \frac{\pi}{4}]$. After applying single Oracle operation $O$, the iteration states shown in (1) are transferred into the Oracle ones

$$|\psi_k^O\rangle \equiv \frac{\cos[(2(k-1)+1)\theta/2]}{\sqrt{2^n}\cos(\theta/2)} \sum_{x\in f^{-1}(0)} |x\rangle - \frac{\sin[(2(k-1)+1)\theta/2]}{\sqrt{2^n}\sin(\theta/2)} \sum_{x\in f^{-1}(1)} |x\rangle. \quad (2)$$

The states in (1) and (2) are of the form

$$|\psi_2\rangle \equiv a \sum_{x\in f^{-1}(0)} |x\rangle + b \sum_{x\in f^{-1}(1)} |x\rangle, \quad (3)$$

where $a, b \in \mathbb{R}$ and $a^2 |f^{-1}(0)| + b^2 |f^{-1}(1)| = 1$. The above states will be referred to as $n$-qubit "*real 2-value states*". It is clear that the REWS's is of the real *2*-value states. In this work, we will investigate the multipartite entanglement features of the real *2*-value states by means of the separable degree and the entanglement measure shown in [11]. Subsequently, we will give the quantitative and qualitative descriptions of the entanglement dynamics in Grover's search algorithm.

The paper is organized as follows. In Sec. II we will review the definitions of separable degree, and maximum Schmidt number of *n*-qubit pure states. Moreover, we will study the multipartite entanglement features of the real *2*-value states by using the separable degree and maximum Schmidt number. In Sec. III we will firstly analyze the entanglement properties of the iteration and Oracle states that occur in Grover's search algorithm. Then we will give the quantitative and qualitative descriptions of the entanglement dynamics in Grover's search algorithm. We will summarize the results in Sec. IV.

## II. Multipartite entanglement and real *2*-value states
### A. Separable degree

In this section we will review the definitions of separable degree [10]. Suppose $|\psi\rangle$ is a pure state of $n$ qubits. If $|\psi\rangle$ can be written as a tensor product of pure state(s) of $k$ individual subsystem(s), then $|\psi\rangle$ is called *k-separable*. Denote by $S_k$ the set of *k*-separable states, then it is obvious that $S_n \subset ... \subset S_2 \subset S_1$. If $|\psi\rangle \in S_k - S_{k+1}$ where $k \in \{1, 2, ..., n-1\}$, then $\delta(|\psi\rangle) \equiv k$ is called by the *separable degree* of $|\psi\rangle$. Moreover, $\delta(|\psi\rangle) = n$ if and only if $|\psi\rangle \in S_n$. If $\delta(|\psi\rangle) = 1$, then $|\psi^n\rangle$ is called *fully entangled*. If $\delta(|\psi\rangle) = n$, then $|\psi^n\rangle$ is *fully separable*. If $n \geq 3$ and $\delta(|\psi\rangle) \in \{2, 3, ..., n-1\}$, then $|\psi\rangle$ is called by a *partially*

*separable state*. In the following, we only consider $n \geq 3$ to be always satisfied.

**B. Qualitative analysis**

In this section we will study the multipartite entanglement features of the real 2-value states by using the separable degree. If $|f^{-1}(1)| \in \{0, 2^n\}$, $|\psi_2\rangle = \pm \frac{1}{\sqrt{2^n}} \sum_{x=0}^{2^n-1} |x\rangle$ are seen as the uniform superposition state which is fully separable. If $|f^{-1}(1)| \notin \{0, 2^n\}$ and $a = 0$, we get

$$|\psi_2\rangle = \frac{1}{\sqrt{|f^{-1}(1)|}} \sum_{x \in f^{-1}(1)} |x\rangle. \tag{4}$$

*Lemma 1*. The state in (4) is fully separable if and only if it satisfies one of the following conditions: (i) $|f^{-1}(1)| = 1$; (ii) there exist $m \in \{1, 2, ..., n-1\}$ and $\omega \in \{0,1\}^{n-m}$ such that

$$|f^{-1}(1)| = 2^m \text{ and } |\psi_2\rangle = \frac{1}{\sqrt{2^m}} \sum_{x=0}^{2^m-1} |x\rangle \otimes |\omega\rangle.$$

Poof: (if) It is obvious. (only if) Since the state in (4) is fully separable, it can be written into the form $\bigotimes_{i=1}^{n}(\alpha_i|0\rangle + \beta_i|1\rangle)$ where $\alpha_i, \beta_i \in \mathbb{R}$. $|f^{-1}(1)| \notin \{0, 2^n\}$ implies that for any $i$ either $\alpha_i = \beta_i$ or $\alpha_i \cdot \beta_i = 0$. Denote by $m$ the number of $\alpha_i = \beta_i$. Thus $m \in \{1, 2, ..., n-1\}$ and $|f^{-1}(1)| = 2^m$. □

Obviously, for $|f^{-1}(1)| \notin \{0, 2^n\}$ and $b = 0$ we can also get similar result. If $a = -b$, we can give

$$|\psi_2\rangle = \frac{1}{\sqrt{2^n}} \sum_{x=0}^{2^n-1} (-1)^{f(x)} |x\rangle \tag{5}$$

which is of the REWS's. [8] and [10] have qualitatively analyzed the multipartite entanglement of the above states in detail.

*Lemma 2*. Suppose $|f^{-1}(1)| \notin \{0, 2^n\}$, $a \neq -b$ and $a \cdot b \neq 0$. If $|\psi_2\rangle$ is 2-separable, it is of the following form

$$|\psi_2\rangle = \left[\frac{1}{\sqrt{2}}(|0\rangle + |1\rangle)\right]^{\otimes k} \otimes \sqrt{2^k} \left(a \sum_{x \in S} |x\rangle + b \sum_{x \in T} |x\rangle\right), \tag{6}$$

where $k \in \{1, 2, ..., n-1\}$, $S \cup T = \{0,1\}^{n-k}$, $S \cap T = \Phi$ and $|T| = |f^{-1}(1)|/2^k$.

Proof: Since $|\psi_2\rangle$ is 2-separable, there exists $k \in \{1, 2, ..., n-1\}$ such that

$|\psi_2\rangle = \left(\sum_{i=0}^{2^k-1} \alpha_i |i\rangle\right) \otimes \left(\sum_{j=0}^{2^{n-k}-1} \beta_j |j\rangle\right)$. Without loss of generality we suppose $\alpha_i, \beta_j \in \mathbb{R}$. For any $i, j$, it is clear that $\alpha_i \cdot \beta_j \neq 0$ since $a \cdot b \neq 0$. Assume that both $\sum_{i=0}^{2^k-1} \alpha_i |i\rangle$ and $\sum_{j=0}^{2^{n-k}-1} \beta_j |j\rangle$ were not uniform superposition states, which implies that there exist $i_1, i_2 \in \{0, 1, \ldots, 2^k-1\}$, $j_1, j_2 \in \{0, 1, \ldots, 2^{n-k}-1\}$ such that $\alpha_{i_1} \neq \alpha_{i_2}$ and $\beta_{j_1} \neq \beta_{j_2}$. Thus $\alpha_{i_1}\beta_{j_1} \neq \alpha_{i_1}\beta_{j_2}$ and $\alpha_{i_2}\beta_{j_1} \neq \alpha_{i_2}\beta_{j_2}$. Since $|\psi_2\rangle$ is the real 2-value state, $\alpha_{i_1}\beta_{j_1} = \alpha_{i_2}\beta_{j_2}$ and $\alpha_{i_1}\beta_{j_2} = \alpha_{i_2}\beta_{j_1}$ which implies $\beta_{j_1} = -\beta_{j_2}$. Thus $\alpha_{i_1}\beta_{j_1} = -\alpha_{i_1}\beta_{j_2}$ which is a contradiction with $a \neq -b$. □

According to the above lemma, for any $|f^{-1}(1)| \notin \{0, 2^n\}$ the number of 2-separable states is given by $nB(2^{n-1}, |f^{-1}(1)|/2)$. Therefore, if $|f^{-1}(1)| < 2^{n/2}$, the faction of 2-separable states is obtained by $\lim_{n \to \infty} \frac{nB(2^{n-1}, |f^{-1}(1)|/2)}{B(2^n, |f^{-1}(1)|)} \simeq 0$ where $B$ denotes the binomial coefficient [8]. We can conclude that for any $|f^{-1}(1)| \in (0, 2^{n/2})$ the states in (3) are almost fully entangled. Moreover, according to the above lemma we can also get the following conclusions.

*Theorem 3.* Suppose $|f^{-1}(1)| \notin \{0, 2^n\}$, $a \neq -b$ and $a \cdot b \neq 0$. Then

(i) $|\psi_2\rangle$ is fully separable if and only if $|f^{-1}(1)| = 2^{n-1}$ and $|\psi_2\rangle$ is one of the forms

$\left[\frac{1}{\sqrt{2}}(|0\rangle + |1\rangle)\right]^{\otimes(n-1)} \otimes \sqrt{2^{n-1}}(a|0\rangle + b|1\rangle)$ and $\left[\frac{1}{\sqrt{2}}(|0\rangle + |1\rangle)\right]^{\otimes(n-1)} \otimes \sqrt{2^{n-1}}(b|0\rangle + a|1\rangle)$

.

(ii) If $|f^{-1}(1)|$ is odd, $|\psi_2\rangle$ is fully entangled.

(iii) If $|f^{-1}(1)| = 2^q(2p+1)$ where $p \in \mathbb{N}$ and $q \in \mathbb{Z}^+$, $|\psi_2\rangle$ is fully entangled or $k$-separable with $k \geq 2$. If $|\psi_2\rangle$ is $k$-separable, then $k \leq q+1$ and it is of the form $\left[\frac{1}{\sqrt{2}}(|0\rangle + |1\rangle)\right]^{\otimes k-1} \otimes \sqrt{2^{k-1}}\left(a\sum_{x \in S}|x\rangle + b\sum_{x \in T}|x\rangle\right)$, where $S \cup T = \{0,1\}^{n-k+1}$,

$S \cap T = \Phi$ and $|T| = 2^{q-k+1}(2p+1)$.

### C. Quantitative analysis

In this section we will use the measure of entanglement shown in [11] to quantify the multipartite entanglement of the real 2-value states. Let $\chi(|\psi\rangle)$ be the maximum Schmidt number of the n-qubit pure state $|\psi\rangle$ over all possible bipartite splitting $A:B$ of n qubits, i.e.,

$$\chi(|\psi\rangle) \equiv \max_A rank\left[Tr_B(|\psi\rangle\langle\psi|)\right], \quad (7)$$

which fulfills the following properties: (i) $\chi(|\psi\rangle) \in \{1, 2, ..., 2^{\lfloor n/2 \rfloor}\}$, with $\chi(|\psi\rangle) = 1$ if and only if it is fully separable; (ii) $\chi(|\psi\rangle \otimes |\psi'\rangle) = \chi(|\psi\rangle) \cdot \chi(|\psi'\rangle)$; (iii) $\chi(|\psi\rangle)$ decreases under LOCC or SLOCC. Then the entanglement measure of $|\psi\rangle$ is defined by $E_\chi(|\psi\rangle) \equiv \log_2(\chi(|\psi\rangle))$ which is an entanglement monotone [12] under LOCC or SLOCC. In this section, we quantitatively investigate multipartite entanglement properties of the real 2-value states $|\psi_2\rangle$ in (3) by directly using the maximum Schmidt number $\chi(|\psi_2\rangle)$. According to the conclusions in Sec. B and the properties of the maximum Schmidt number, we can get several results as follows.

*Thereom 4.* If $|f^{-1}(1)| \in \{0, 2^n\}$, then $\chi = 1$.

*Thereom 5.* Suppose $|f^{-1}(1)| \notin \{0, 2^n\}$ and $a = 0$. Then

(i) If $\chi = 1$, there exists $m \in \{0, 1, ..., n\}$ such that $|f^{-1}(1)| = 2^m$.

(ii) If $|f^{-1}(1)| = 1$, $\chi = 1$.

(iii) If $|f^{-1}(1)| \in \{2^m | m \in \{1, 2, ..., n-1\}\}$, $\chi \in \{1, 2, ..., \min(|f^{-1}(1)|, 2^{\lfloor n/2 \rfloor})\}$.

(iv) If $|f^{-1}(1)| \notin \{2^m | m \in \{0, 1, ..., n-1\}\}$, $\chi \in \{2, 3, ..., \min(|f^{-1}(1)|, 2^{\lfloor n/2 \rfloor})\}$.

Obviously, for $|f^{-1}(1)| \notin \{0, 2^n\}$ and $b = 0$ we can also get similar results.

*Thereom 6.* Suppose $a = -b$. Then

(i) If $\chi = 1$, $|f^{-1}(1)| = |f^{-1}(0)| = 2^{n-1}$.

(ii) If $|f^{-1}(1)| = 2^{n-1}$, $\chi \in \{1, 2, ..., 2^{\lfloor n/2 \rfloor}\}$.

(iii) If $|f^{-1}(1)| \neq 2^{n-1}$, $\chi \in \{2,3,...,\min(|f^{-1}(1)|+1, 2^{\lfloor n/2 \rfloor})\}$.

(iv) If $|f^{-1}(1)| = 1$, $\chi = 2$.

*Thereom 7.* Suppose $|f^{-1}(1)| \notin \{0, 2^n\}$, $a \neq -b$ and $a \cdot b \neq 0$. Then

(i) If $\chi = 1$, $|f^{-1}(1)| = |f^{-1}(0)| = 2^{n-1}$.

(ii) If $|f^{-1}(1)| = 2^{n-1}$, $\chi \in \{1, 2, ..., 2^{\lfloor n/2 \rfloor}\}$.

(iii) If $|f^{-1}(1)| \neq 2^{n-1}$, $\chi \in \{2, 3, ..., \min(|f^{-1}(1)|+1, 2^{\lfloor n/2 \rfloor})\}$.

(iv) If $|f^{-1}(1)| = 1$, $\chi = 2$.

Now, we only prove the results (iv) in thereom 6 and (iv) in thereom 7.

Proof: Without loss of generality we suppose $|\psi_2\rangle = a \sum_{x=1}^{2^n - 1} |x\rangle + b|0\rangle$ with $a \cdot b \neq 0$. We can obtain the density matrix $|\psi_2\rangle\langle\psi_2|$, i.e.,

$$\begin{bmatrix} b^2 & ab & ab & \cdots & ab \\ ab & a^2 & a^2 & \cdots & a^2 \\ ab & a^2 & a^2 & \cdots & a^2 \\ \vdots & \vdots & \vdots & \ddots & \vdots \\ ab & a^2 & a^2 & \cdots & a^2 \end{bmatrix}_{2^n \times 2^n}.$$

The reduced density matrix $Tr_B(|\psi_2\rangle\langle\psi_2|)$ for any bipartite splitting $A:B$, where $m$ qubits are in $A$ and $n-m$ in $B$, is given

$$\begin{bmatrix} b^2 + (2^{n-m}-1)a^2 & ab + (2^{n-m}-1)a^2 & ab + (2^{n-m}-1)a^2 & \cdots & ab + (2^{n-m}-1)a^2 \\ ab + (2^{n-m}-1)a^2 & 2^{n-m}a^2 & 2^{n-m}a^2 & \cdots & 2^{n-m}a^2 \\ ab + (2^{n-m}-1)a^2 & 2^{n-m}a^2 & 2^{n-m}a^2 & \cdots & 2^{n-m}a^2 \\ \vdots & \vdots & \vdots & \ddots & \vdots \\ ab + (2^{n-m}-1)a^2 & 2^{n-m}a^2 & 2^{n-m}a^2 & \cdots & 2^{n-m}a^2 \end{bmatrix}_{2^m \times 2^m}.$$

Since $a \cdot b \neq 0$, $rank[Tr_B(|\psi_2\rangle\langle\psi_2|)] = 2$. □

In the following section we will investigate the entanglement properties of the iteration and Oracle states that occur in Grover's search algorithm according to the conclusions in this section. Subsequently, we will give the quantitative and qualitative descriptions of the entanglement dynamics.

**III. Entanglement dynamics of Grover's search algorithm**

In this paper, we adopt the quantum circuit shown in [1] to implement Grover's search algorithm. We only consider the multipartite entanglement of the first $n$ qubits in Grover's search algorithm. The initial state is $|0\rangle^{\otimes n}$ which is fully separable and $\chi=1$. Subsequently, the state $|\psi_0^G\rangle$ is achieved by applying the Hadamard transformation $H^{\otimes n}$ in the first $n$ qubits, where $H=\frac{1}{\sqrt{2}}\begin{bmatrix}1 & 1 \\ 1 & -1\end{bmatrix}$. It is obvious that $|\psi_0^G\rangle$ is also fully separable and $\chi=1$. Then $R\equiv\left\lceil\frac{\pi-\theta}{2\cdot\theta}-\frac{1}{2}\right\rceil$ Grover iterations $G\equiv PO$, where $P$ is the inversion about mean operation $2|\psi_0^G\rangle\langle\psi_0^G|-I$, are performed. The iteration and Oracle states achieved by applying Grover iterations are respectively shown in the forms (1) and (2). Finally, we can get one of the desired solution(s) with high probability $\varepsilon\in\left[\frac{1}{2},1\right]$ by measuring the first $n$ qubits. Thus the states draw a picture of the dynamics of Grover's search algorithm as follows.

$$|0\rangle^{\otimes n}\xrightarrow{H^{\otimes n}}|\psi_0^G\rangle\xrightarrow{O}|\psi_1^O\rangle\xrightarrow{P}|\psi_1^G\rangle\xrightarrow{O}\cdots\xrightarrow{O}|\psi_R^O\rangle\xrightarrow{P}|\psi_R^G\rangle \qquad (8)$$

If $|f^{-1}(1)|\geq 2^{n-1}$, $R=0$. Thus one of the desired solutions is given with probability at least $\frac{1}{2}$. Note that $|f^{-1}(1)|$ can be approximately obtained by quantum counting algorithm [1]. In the following, we only consider the condition $|f^{-1}(1)|\in\{1,2,...,2^{n-1}-1\}$ to be always fulfilled. We will study the multipartite entanglement description of the dynamics of Grover's search algorithm for $|f^{-1}(1)|<2^{n-1}$.

*Lemma 8.* For any $k\in\{1,2,...,R\}$, $\frac{\cos[(2k+1)\theta/2]}{\sqrt{2^n}\cos(\theta/2)}\neq\frac{\sin[(2k+1)\theta/2]}{\sqrt{2^n}\sin(\theta/2)}$.

Poof: Assume that $\frac{\cos[(2k+1)\theta/2]}{\sqrt{2^n}\cos(\theta/2)}=\frac{\sin[(2k+1)\theta/2]}{\sqrt{2^n}\sin(\theta/2)}$, which implies $\sin(k\theta)=0$.

Thus it would exist $r\in\mathbb{Z}^+$ such that $\theta=r\cdot\pi/k$. Since $R=\left\lceil\frac{\pi-\theta}{2\cdot\theta}-\frac{1}{2}\right\rceil\leq\frac{\pi-\theta}{2\cdot\theta}+\frac{1}{2}$, we get $R\leq\frac{\pi}{2\cdot\theta}=\frac{\pi}{2\cdot r\cdot\pi/k}=\frac{k}{2\cdot r}$ which is a contradiction with $k\leq R$. $\square$

**A. Qualitative analysis**

According to the conclusions in Sec. II B, we can get the following results.

*Thereom 9.* Suppose $|f^{-1}(1)|$ is odd. Then

(i) All of $|\psi_1^O\rangle$, $|\psi_1^G\rangle$,..., $|\psi_{R-1}^O\rangle$, $|\psi_{R-1}^G\rangle$ and $|\psi_R^O\rangle$ are fully entangled.

(ii) If $\cos[(2R+1)\theta/2] \neq 0$, $|\psi_R^G\rangle$ is fully entangled. Otherwise, $|\psi_R^G\rangle$ is fully entangled or partially separable, but it is not fully separable.

*Thereom 10.* Suppose $|f^{-1}(1)| = 2^q(2p+1)$ where $p \in \mathbb{N}$ and $q \in \mathbb{Z}^+$. Then

(i) Suppose $|\psi_1^O\rangle$ is fully entangled. Then all of $|\psi_1^G\rangle$,..., $|\psi_{R-1}^O\rangle$, $|\psi_{R-1}^G\rangle$ and $|\psi_R^O\rangle$ are fully entangled. If $\cos[(2R+1)\theta/2] \neq 0$, $|\psi_R^G\rangle$ is fully entangled.

(ii) Suppose $|\psi_1^O\rangle$ is of the form

$$\left[\frac{1}{\sqrt{2}}(|0\rangle+|1\rangle)\right]^{\otimes k-1} \otimes \sqrt{2^{k-1}}\left(\sum_{x \in S}|x\rangle - \sum_{x \in T}|x\rangle\right) \quad (9)$$

where $k \in \{2,...,q+1\}$, $S \cup T = \{0,1\}^{n-k+1}$, $S \cap T = \Phi$ and $|T| = 2^{q-k+1}(2p+1)$. Then all of $|\psi_1^G\rangle, |\psi_2^O\rangle,..., |\psi_{R-1}^G\rangle$ and $|\psi_R^O\rangle$ are k-separable. Moreover, $\forall i \in \{1,2,...,R-1\}$, $|\psi_i^G\rangle$ is of the form

$$\left[\frac{1}{\sqrt{2}}(|0\rangle+|1\rangle)\right]^{\otimes k-1} \otimes \sqrt{2^{k-1}}\left(\frac{\cos[(2i+1)\theta/2]}{\sqrt{2^n}\cos(\theta/2)}\sum_{x \in S}|x\rangle + \frac{\sin[(2i+1)\theta/2]}{\sqrt{2^n}\sin(\theta/2)}\sum_{x \in T}|x\rangle\right) \quad \text{and}$$

$|\psi_{i+1}^O\rangle$ is of the form

$$\left[\frac{1}{\sqrt{2}}(|0\rangle+|1\rangle)\right]^{\otimes k-1} \otimes \sqrt{2^{k-1}}\left(\frac{\cos[(2i+1)\theta/2]}{\sqrt{2^n}\cos(\theta/2)}\sum_{x \in S}|x\rangle - \frac{\sin[(2i+1)\theta/2]}{\sqrt{2^n}\sin(\theta/2)}\sum_{x \in T}|x\rangle\right) \quad .$$

If $\cos[(2R+1)\theta/2] \neq 0$, $|\psi_R^G\rangle$ is k-separable. Otherwise, $|\psi_R^G\rangle$ is j-separable with $j \geq k$.

(iii) $|\psi_R^G\rangle$ is fully separable if and only if $\cos[(2R+1)\theta/2]=0$, $p=0$ and $|\psi_1^O\rangle$ is q+1-separable.

Note that the multipartite entanglement of the real equally weighted states is very complex if $|f^{-1}(1)| = 2^q(2p+1) \geq 2^{n/2}$ and $p \geq 1$ [10]. Thus $|\psi_1^O\rangle$ with $\delta(|\psi_1^O\rangle) = k \geq 2$ might have different form with the one shown in (9), which implies that all separable degrees of $|\psi_1^G\rangle, |\psi_2^O\rangle,..., |\psi_{R-1}^G\rangle$ and $|\psi_R^O\rangle$ might be smaller than k though $|\psi_1^O\rangle$ is k-separable.

**B. Quantitative analysis**

According to the conclusions in Sec. II B and C, we can get several results as follows.

*Thereom 11.* Suppose $|f^{-1}(1)|$ is odd. Then

(i)
$$\chi(|\psi_1^O\rangle), \chi(|\psi_1^G\rangle), ..., \chi(|\psi_{R-1}^O\rangle), \chi(|\psi_{R-1}^G\rangle), \chi(|\psi_R^O\rangle) \in \{2, 3, ..., \min(|f^{-1}(1)|+1, 2^{\lfloor n/2 \rfloor})\}.$$

In particular, $\chi(|\psi_1^O\rangle) = \chi(|\psi_1^G\rangle) = ... = \chi(|\psi_{R-1}^O\rangle) = \chi(|\psi_{R-1}^G\rangle) = \chi(|\psi_R^O\rangle) = 2$ if $|f^{-1}(1)| = 1$.

(ii) $\chi(|\psi_R^G\rangle) = 1$ if and only if $\cos[(2R+1)\theta/2] = 0$ and $|f^{-1}(1)| = 1$.

*Thereom 12.* Suppose $|f^{-1}(1)| = 2^q(2p+1) < 2^{n/2}$ where $p \in \mathbb{N}$ and $q \in \mathbb{Z}^+$. Then

(i) Suppose $|\psi_1^O\rangle$ is fully entangled. Then $\chi(|\psi_1^O\rangle), \chi(|\psi_1^G\rangle), ..., \chi(|\psi_{R-1}^O\rangle), \chi(|\psi_{R-1}^G\rangle)$,
$\chi(|\psi_R^O\rangle) \in \{2, 3, ..., \min(|f^{-1}(1)|+1, 2^{\lfloor n/2 \rfloor})\}$. If $\cos[(2R+1)\theta/2] \neq 0$, $\chi(|\psi_R^G\rangle)$
$\in \{2, 3, ..., \min(|f^{-1}(1)|+1, 2^{\lfloor n/2 \rfloor})\}$. Otherwise, $\chi(|\psi_R^G\rangle) \in \{2, 3, ..., \min(|f^{-1}(1)|, 2^{\lfloor n/2 \rfloor})\}$.

(ii) Suppose $|\psi_1^O\rangle = \left[\frac{1}{\sqrt{2}}(|0\rangle+|1\rangle)\right]^{\otimes k-1} \otimes \sqrt{2^{k-1}}\left(\sum_{x \in S}|x\rangle - \sum_{x \in T}|x\rangle\right)$ where
$k \in \{2, ..., q+1\}$, $S \cup T = \{0,1\}^{n-k+1}$, $S \cap T = \Phi$, $|T| = 2^{q-k+1}(2p+1)$ and
$\sqrt{2^{k-1}}\left(\sum_{x \in S}|x\rangle - \sum_{x \in T}|x\rangle\right)$ is fully entangled. Then

$$\chi(|\psi_1^O\rangle), \chi(|\psi_1^G\rangle), ..., \chi(|\psi_{R-1}^O\rangle), \chi(|\psi_{R-1}^G\rangle), \chi(|\psi_R^O\rangle) \in \{2, 3, ..., \min(|T|+1, 2^{\lfloor (n-k+1)/2 \rfloor})\}.$$

If $\cos[(2R+1)\theta/2] = 0$, $p = 0$ and $|\psi_1^O\rangle$ is q+1-separable, $\chi(|\psi_1^O\rangle) = \chi(|\psi_1^G\rangle) = ...$
$= \chi(|\psi_{R-1}^O\rangle) = \chi(|\psi_{R-1}^G\rangle) = \chi(|\psi_R^O\rangle) = 2$. Moreover, if $\cos[(2R+1)\theta/2] \neq 0$, $\chi(|\psi_R^G\rangle)$
$\in \{2, 3, ..., \min(|T|+1, 2^{\lfloor (n-k+1)/2 \rfloor})\}$. Otherwise, $\chi(|\psi_R^G\rangle) \in \{1, 2, 3, ..., \min(|T|, 2^{\lfloor (n-k+1)/2 \rfloor})\}$.

(iii) $\chi(|\psi_R^G\rangle) = 1$ if and only if $\cos[(2R+1)\theta/2] = 0$, $p = 0$ and $|\psi_1^O\rangle$ is q+1-separable.

## IV. Conclusions

As shown in Tab. 1, we give the quantitative and qualitative descriptions of the entanglement dynamics of Grover's search algorithm. Suppose $|f^{-1}(1)| \in (0, 2^{n/2})$. For most instances, it is clear that $\cos[(2R+1)\theta/2] \neq 0$. Thus the separable degrees of $|\psi_1^O\rangle$, $|\psi_1^G\rangle$,..., $|\psi_R^O\rangle$ and $|\psi_R^G\rangle$ are invariable. Since for most instances, $|\psi_1^O\rangle$ is fully entangled [8, 10] and then $|\psi_1^G\rangle$,..., $|\psi_R^O\rangle$ and $|\psi_R^G\rangle$ are also fully entangled. Thus the dynamics of Grover' search algorithm is almost "filled" by the fully entangled states. And the maximum Schmidt numbers of $|\psi_1^O\rangle$, $|\psi_1^G\rangle$,..., $|\psi_R^O\rangle$ and $|\psi_R^G\rangle$ have the same range. In particular, if $|f^{-1}(1)| = 2^{\delta(|\psi_1^O\rangle)-1}$, all maximum Schmidt numbers of $|\psi_1^O\rangle$, $|\psi_1^G\rangle$,..., $|\psi_R^O\rangle$ and $|\psi_R^G\rangle$ are equal to *2*.

Table 1. The quantitative and qualitative descriptions of the entanglement dynamics in Grover's search algorithm, where $A = \{2,3,...,\min(|f^{-1}(1)|+1, 2^{\lfloor n/2 \rfloor})\}$, $B = \{2,3,...,\min(|f^{-1}(1)|/2^{k-1}+1, 2^{\lfloor (n-k+1)/2 \rfloor})\}$, $A' = \{2,3,...,\min(|f^{-1}(1)|, 2^{\lfloor n/2 \rfloor})\}$ and $B' = \{2,3,...,\min(|f^{-1}(1)|/2^{k-1}, 2^{\lfloor (n-k+1)/2 \rfloor})\}$.

| $|f^{-1}(1)|$ | | $|\psi_1^O\rangle$ | | $|\psi_1^G\rangle$ | | ... | $|\psi_{R-1}^O\rangle$ | | $|\psi_{R-1}^G\rangle$ | | $|\psi_R^O\rangle$ | | $|\psi_R^G\rangle$ | | | |
|---|---|---|---|---|---|---|---|---|---|---|---|---|---|---|---|---|
| | | | | | | | | | | | | | $\cos[(2R+1)\theta/2] \neq 0$ | | $\cos[(2R+1)\theta/2] = 0$ | |
| $\in \{1,2,...,2^{n-1}-1\}$ | | $\delta$ | $\chi$ | $\delta$ | $\chi$ | ... $\delta$ | $\chi$ | $\delta$ | $\chi$ | $\delta$ | $\chi$ | $\delta$ | $\chi$ | $\delta$ | $\chi$ |
| 1 | | 1 | 2 | 1 | 2 | ... 1 | 2 | 1 | 2 | 1 | 2 | 1 | 2 | n | 1 |
| $2p+1$, $p \in \mathbb{Z}^+$ | | 1 | $\in A$ | 1 | $\in A$ | ... 1 | $\in A$ | 1 | $\in A$ | 1 | $\in A$ | 1 | $\in A$ | $\in \{1,2,...,n-1\}$ | $\in A'$ |
| $2^q$, $q \in \mathbb{Z}^+$ | 1 | | $\in A$ | 1 | $\in A$ | ... 1 | $\in A$ | 1 | $\in A$ | 1 | $\in A$ | 1 | $\in A$ | $\in \{1,2,...,n-1\}$ | $\in A'$ |
| | $k \in \{2,3,...,q\}$ | $\in B$ | k | $\in B$ | ... k | $\in B$ | k | $\in B$ | k | $\in B$ | k | $\in B$ | $\in \{k,k+1,...,n-1\}$ | $\in B'$ |
| | q+1 | 2 | q+1 | 2 | ... q+1 | 2 | q+1 | 2 | q+1 | 2 | q+1 | 2 | n | 1 |
| $2^q(2p+1) < 2^{n/2}$, $p,q \in \mathbb{Z}^+$ | 1 | | $\in A$ | 1 | $\in A$ | ... 1 | $\in A$ | 1 | $\in A$ | 1 | $\in A$ | 1 | $\in A$ | $\in \{1,2,...,n-1\}$ | $\in A'$ |
| | $k \in \{2,3,...,q+1\}$ | $\in B$ | k | $\in B$ | ... k | $\in B$ | k | $\in B$ | k | $\in B$ | k | $\in B$ | $\in \{k,k+1,...,n-1\}$ | $\in B'$ |

**ACKNOWLEDGMENTS**

This work was financially supported by the National Natural Science Foundation of China under Grant No. 61170178.